# Expertise Indices: Variants, Modifications, Advancements, and Computational Tools in R

Abhirup Nandy [1, #], Nilabhra R. Das [2, #]

1. Shri Ram College of Commerce, University of Delhi, New Delhi, India
2. Mater Research Institute, The University of Queensland, Brisbane, Australia
\# These authors contributed equally.

# Abstract

In the academic landscape, scientific research has been primarily conducted through research institutions, which requires a massive influx of funds from various sources. Presently, these funding bodies have been moving from trust-based funding to performance- based evaluation systems for granting funds to the research bodies. This has led to the rise in popularity of various indices or statistics that measure institutional research strength or expertise. Institutional research expertise usually focuses on publication volume and its impact measured using the widely used $h$- and $g$-indices. However, these indices fail to capture the thematic expertise of research for institutions. To address this gap, two new expertise indicators, namely the $x$-index, the $x_d$-index, and bias-adjusted variants, the field-normalised $x_d$-index, and the fractional $x_d$-index, were introduced recently. Additionally, we propose two new variants, the category-adjusted $x$-index and the inverse variance weighted $x_d$-index, which further account for resolvable bias, and a novel statistic, the $x_o$-index, which acts as a measure of the overall research expertise. While several packages that calculate the traditional $h$- and $g$-indices exist, these novel expertise indices are yet to be included in such existing packages. The 'xxdi' R package provides simple functions that implement these expertise indices and their variants, enabling their utilisation by the wider research community. A stable version of the package is available on CRAN (https://doi.org/10.32614/CRAN.package.xxdi) and an in-development version on GitHub (https://github.com/nilabhrardas/xxdi).

**Keywords:** bibliometric indices, collaboration planning, expertise indices, institutional research impact, R, xxdi

# Introduction

The research impact of institutions and individual authors plays a crucial role in securing funding for different research projects. Hence, the evaluation of the research strength of these institutions have long been of importance to multiple stakeholders, at various levels of governance. The assessment of research strength of institutions is consequential in providing information on the reliability of the research conducted and the strength of academic faculty. This would help to support critical decision-making processes among public administrators, students surveying study opportunities, researchers considering better career options, and other stakeholders. Funding for research (primarily derived from public funds) is limited and may become inadequate provisions when allocated to all (for example, all institutions within a country) in equal proportions.

In recent years, funding bodies have adopted a 'performance-based' approach to allocate funds (Lathabai et al., 2021b), resulting in the need for assessing and ranking academic institutions based on research strength or expertise. To gauge and compare research strength of institutions, several metrics have been devised (Franceschini & Maisano, 2011; Lazaridis, 2010). Amongst these, Hirsh's $h$-index (Hirsch, 2005) and Egghe's $g$-index (Egghe, 2006) are the most commonly used measures of research impact, and have been widely applied to both authors and institutions. Since their inception, a large amount of literature has utilised these metrics to assess the research impact of institutions.

However, at the institution level, evaluating research impact solely based on publication volume and citation counts may lead to inaccurate estimates. Multiple studies have raised concerns over the use of the $h$- and g-indices as the main metrics for evaluating research activity. For instance, the h-index does not correctly reflect scientific impact, as authors with Nobel prizes have traditionally had lower h-indices compared to those without (Hirsch, 2005). The $h$-index is based on research output, not accounting for the disparity in publication criteria across different research fields. Furthermore, both of these indices ignore the contribution from articles with citation numbers outside of the core works, although these works might be steadily gaining attention (Ye & Rousseau, 2010). On the other hand, Huang and Lin (2011) highlighted the inconsistency of the h-index in assessing institutional performance using 20-year citations records on Web of Science (WoS). Their study showed that three different affiliation counting methods (whole, straight, and fractional counting) resulted in notably different $h$-index values, consequently affecting institutional rankings.

Since these indices were proposed as measures of research impact of individual authors, they fail when it comes to evaluating the true research impact and capacity of institutions.

Most importantly, scholarly institutions are typically multidisciplinary but often exhibit greater specialisation in certain research fields over others. As a result, the volume of publication outputs and citation counts are likely to be skewed towards these dominant disciplines, obscuring the institution's broader thematic profile. However, the computation processes of the h- and g-indices do not account for research areas or research fields (Lathabai et al., 2021a; Nandy et al., 2023). For instance, an institution that does not perform satisfactorily overall in institutional rankings, may be a leading institute in some niche research fields. Such fine-grained nuances are overlooked when using the $h$- or the g-indices. Thus, a framework was required to determine the core competency and thematic research strengths of institutions as these two factors together can effectively represent an institution's research strength at a detailed level of granularity (López-Illescas et al., 2011). This limitation has motivated the modification of the h- and g-indices to develop metrics that explicitly incorporate intra-institutional multidisciplinary structure. Two new expertise indicators have been proposed recently, the expertise index ($x$-index) (Lathabai et al., 2021b) and the expertise diversity index ($x_d$-index) (Nandy et al., 2023). These indicators have been designed to quantify the thematic expertise and diversity of research output, making them particularly well suited for application to scholarly institutions and other research bodies.

## $x$-index: core competency within a discipline

The $x$-index is a measure of disciplinary expertise depth based on fine-grained thematic areas within disciplines (Lathabai et al., 2021b). These fine-grained themes are usually captured by keywords, which may be author provided or indexed keywords from scholarly databases (such as the WoS database). To compute the $x$-index, thematic areas are ranked by citation strength and evaluated using a Citation-to-Rank Ratio (CRR). A threshold of CRR ≥ 1 is used to identify the top impactful areas.

The $x$-index is best suited to discipline-specific departments or schools within scholarly bodies. A department/school has an $x$-index of $x$, i.e., a research expertise depth of $x$ if it has published in at least $x$ fine thematic areas and received at least $x$ citations in each of those areas (Lathabai et al., 2021b). These top $x$ thematic areas represent the department's core competency domains. In simpler terms, a high $x$-index value signifies a deeper research strength across research sub-topics within a discipline.

# Bias-aware variant of the $x$-index

The evolution of scientific inquiry, followed by shifts in research methodology, have led to the emergence of fields such as genetics (including fields such as biology, chemistry, mathematics, medical sciences, etc.), artificial intelligence (computer science, physics, mathematics, etc. as well as applied fields such as medicine and biology), and quantum mechanics (physics, mathematics, computer science, electronics, etc.). These disciplines integrate multiple areas of science, thus driving the establishment of multidisciplinary departments dedicated to the pursuit of research in these disciplines. However, when applied to such departments, the $x$-index is affected by overlap bias which arises from the multidisciplinary nature of the research conducted and/or the collaborative and interdisciplinary aspects of these research fields.

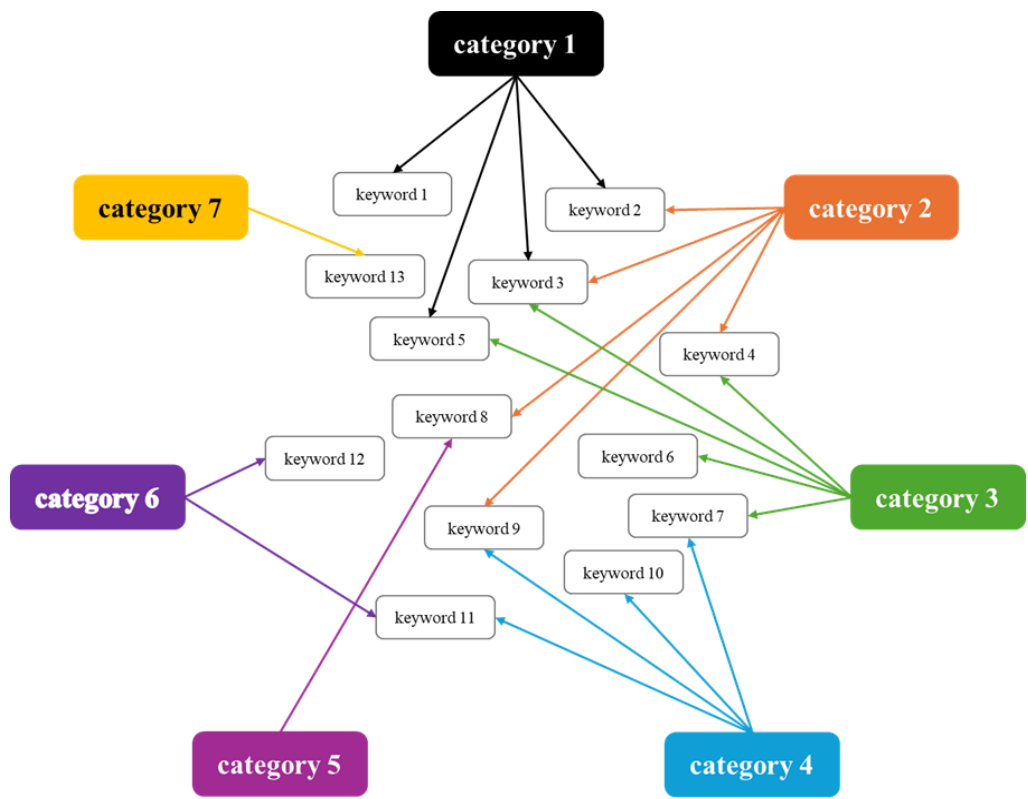

**Figure 1.** Overlap of keywords belonging to different categories.

While the $x$-index utilises low-level publication 'keywords' and ranks them by citation counts, publication databases (such as WoS) also include higher-level 'categories' which represent broader thematic areas, that is, several nested keywords form a single category. Consequently, several keywords are highly likely to be members of the sets of keywords that form two different categories (**Figure 1**), i.e., overlap, resulting in fuzziness in the $x$-index. For instance, the keyword "domain" is prevalent in fields like taxology, ecology, mathematics, and computer science, or the keyword "vector" which is widely prevalent in fields like biology, epidemiology, physics, and mathematics. As illustrated, the same keywords are often used in a multitude of fields to convey different meanings, violating the mutual exclusion restriction assumption in the $x$-index. The presence of overlap bias is likely to attenuate the true research depth, as measured using the $x$-index, in the case of highly multidisciplinary or collaborative departments and conversely, inflate the true research strength for more discipline-specific departments.

Here, we propose the $x_c$-index as an improvement over the traditional $x$-index by adjusting for broader subject categories in the $x$-index computation process. We adapted the $x$-index to derive the $x_c$-index by replacing total citations for unique keywords (thematic strength) with categorically adjusted unique keywords which are a better representation of thematic strength. Thus, the $x_c$-index refines the $x$-index by treating keywords occurring in multiple categories as distinct, thereby providing a more accurate measure of a department's research impact (**Equation 1**).

**Equation 1:**

$$x_c = \begin{cases} r_{ij}, & if\ CRR = 1 \\ min(r_{ij} - 1), & if\ CRR < 1\ at\ r_{ij} \end{cases}$$

where, $r_{ij}$ is the rank assigned to the $i$-th keyword from the $j$-th category based on total citations observed; and CRR is the citation-rank ratio, defined as $\frac{t_{ij}}{r_{ij}}$, where $t_{ij}$ is the total observed citations for the $i$-th keyword belonging to the $j$-th category.

The category adjusted $x$-index, the $x_c$-index, for a department will be $x_c$ if the department has publications in at least $x_c$ thematic areas (identified by category specific keywords) and received at least $x_c$ citations in each of those areas. In other words, keywords are treated as category specific, i.e., if the same keyword appears in multiple categories, each occurrence is considered as a distinct thematic area. For example, a keyword prevalent within three different categories is counted as three unique keywords.

# $x_d$-index: expertise diversity

The $x_d$-index is a measure of the expertise diversity of an institution (Nandy et al., 2023). The $x_d$-index generalises the $x_d$-index to coarse-level thematic areas (such as WoS subject categories), indicating the diversity of impactful research areas or disciplines. An institution has an $x_d$-index of $x_d$ if it publishes in at least $x_d$ subject categories and has received at least $x_d$ citations in each of those categories (Nandy et al., 2023). In simpler terms, the $x_d$-index gives the broad domains with significant scholarly impact within an institution, with high $x_d$-index values reflecting greater thematic diversity, revealing expansive and multidisciplinary portfolios.

## Bias-aware variants of the $x_d$-index

Different research domains attract unequal citation norms (Mendoza, 2021), and collaborative articles may inflate institutional credit (Nandy et al., 2024). To address these concerns, two additional variants were proposed:

### $x_d(f)$-index

A variant of the $x_d$-index that uses fractional citation counts to reduce inflation due to multi-institution collaborations (Nandy et al., 2024). This approach ensures a more balanced attribution of research impact. It reduces the overestimation of institutional expertise that may arise from collaborative intensity rather than intrinsic research strength. In this variant, the citation count for a publication is fractionally allocated by dividing the observed citation count by the number of contributing institutions. These fractional scores are then aggregated at the category level to compute the $x_d$-index, giving the fractional $x_d$-index or $x_d(f)$-index.

### $x_d(FN)$-index

A variant of the $x_d$-index that uses field-normalised citation scores to reduce citation bias across disciplines (Nandy et al., 2024), to minimise disciplinary bias when assessing heterogeneous research portfolios. In this variant, the category-level citation counts are normalised by dividing observed citations by the expected (mean) citation count of the corresponding category, producing inverse mean weighted citation scores. The $x_d$-index is then computed using these normalised scores, giving the field-normalised $x_d$-index or $x_d(FN)$-index.

### IVW $x_d$-index

Additionally, we propose the inverse variance weighted (IVW) $x_d$-index which further improves upon the $x_d(FN)$-index to minimise the impact of disciplinary bias or field bias. While using field-specific mean citation scores is mathematically simple, using a mean-based method, i.e., the $x_d(FN)$-index, still assumes that there is no disciplinary trend. Arithmetic means, treat all fields as homogenous, i.e., it assumes a similar distribution of citations for all research fields being evaluated, or mathematically, equal variances across all fields. This inevitably leads to biased results when citations within a field are more variable than in another field, which may lead to inflation or attenuation of the mean (Mendoza, 2021). Inverse variance weighting minimises this bias by emphasising on the consistency of citation counts within fields. In this variant, the category-level citation counts are weighted by dividing observed citations by the expected variability of citations (variance) of the corresponding category, producing IVW citation scores. The $x_d$-index is then computed using these normalised citation scores, giving the IVW $x_d$-index (**Equation 2**). Similar to the $x_d$-index, the IVW $x_d$-index is defined as the rank (i.e., $r_i$) where the IVW CRR becomes 1, or the preceding rank (i.e., $r_i$ - 1) of the rank where the IVW CRR becomes less than 1.

**Equation 2:**

$$IVW\ x_d = \begin{cases} r_i, & IVW\ CRR = 1 \\ \min(r_i - 1), & IVW\ CRR < 1\ at\ r_i \end{cases}$$

where, $r_i$ is the rank assigned to the $i$-th category based on total citations observed; $IVW\ CRR = v_i^{-1} \times \frac{t_i}{r_i}$, where $t_i$ and $v_i$ are the total observed citations and the variance of the citations (alternatively, the estimated variance from observed citations) for the $i$-th category.

Research field-specific citation variances are best based on the list of all publications in the specific research field till date, i.e., the population variance. However, such a vast scale of data is seldom accessible to all. Variances based on the publications from the institution of interest within the time period of interest may be used as estimates in large samples. In small samples, such as studies involving investigation of an institution with less than 100 publications per research field within the time period of interest, sample variances may not provide robust estimates of the true variability and the use of the traditional $x_d$-index should be considered.

# $xx$-index and $xx_d$-index:

The $x$- and $x_d$-indices may also be used to calculate nested indicator metrics ($xx$-index or $xx_d$-index) to measure expertise depth and breadth at a higher level (such as national research expertise) (Lathabai et al., 2021b; Nandy et al., 2023). For example, we can calculate the $x$-indices for all institutions in a particular country (or geo-political region). Using these $x$-index values as weights for the institutions, we can apply the $h$-index technique again to get the $h$-type $xx$-index for that country (and similarly for the $xx_d$-index where the $x_d$-indices are used as weights for the institutions). These indices help to determine the competency levels for a group of institutions. Such a high-level measure of expertise helps to determine many factors like the growth or diversity of institutions at a regional or national level and are much helpful in policy readiness.

## $x_o$-index: overall expertise, competency-based disciplinary diversity

The $x$- and $x_d$-indices and their different variants measure the thematic depth (primarily departmental) and thematic breadth (primarily institutional) of scholarly bodies respectively. While these indices are good estimates of research competency in fine-grained thematic areas and/or disciplinary diversity in research impact, they fall short when it comes to an overall measure of institutional expertise. To address this gap, we introduce the $x_o$-index, which is a measure of overall institutional research strength, accounting for both the depth and breadth of disciplinary (thematic) expertise. First, the $x$-index is calculated based on all keywords nested within a particular category and the total citations observed for those nested keywords. This is then repeated for all observed categories. Next, the $h$-index technique is applied to those categories ranked based on their respective $x$-indices to give the $x_o$-index (**Equation 3**). Simply, the $x_o$-index is defined as the rank (i.e., $r_i$) where the $x$-rank ratio or XRR (in contrast to the typical CRR) is 1, or the preceding rank (i.e., $r_i$ - 1) of the rank where the IVW CRR becomes less than 1.

**Equation 3:**

$$x_o = \begin{cases} r_i, & XRR = 1 \\ \min(r_i - 1), & XRR < 1 \text{ at } r_i \end{cases}$$

where $r_i$ is the rank assigned to the $i$-th category based on the $x$-index for that category; $XRR = \frac{x_i}{r_i}$, where $x_i$ is the $x$-index calculated using all keywords nested within the $i$-th category and the total citations received by those keywords.

An $x_o$-index of $x$ for an institution suggests research diversity spread across $x$ categories, with a depth of at least $x$ keywords per category.

# xxdi: An R Package for Evaluating Expertise Indices for Research Strength Assessment

While the underlying modifications are mathematically simplistic, computation becomes overly complicated due to the data structure of publicly available databases (such as Web of Science). Several software packages (such as 'agop' (Gagolewski & Cena, 2023) and 'bibliometrix' (Aria & Cuccurullo, 2017) in R) to compute the traditional $h$- and $g$-indices exist. Nonetheless, to our knowledge, these packages are yet to incorporate these novel expertise indicators. Here we introduce an R package, 'xxdi' (Das & Nandy, 2026) which includes user-friendly functions to compute the $x$-, $x_d$-, and $x_o$-indices within the R environment, allowing users to evaluate the depth and breadth of research expertise of research institutions or other scholarly entities. The latest stable version of the package (version 1.3.1) is available for download on CRAN (https://doi.org/10.32614/CRAN.package.xxdi), while an in-development version (version 1.26.4) can be found on GitHub (https://github.com/nilabhrardas/xxdi).

# Discussion

Here we briefly discussed shortcomings of the traditional $h$- and $g$-indices in appropriately evaluating the strength and/or expertise of research bodies in the different disciplines of science (Lathabai et al., 2021b). While the introduction of expertise indices, namely the $x$-, $x_d$-, $xx$-, $xx_d$-, and their different variants have recently enabled analytical advancements in this respect (Lathabai et al., 2021b; Nandy et al., 2024), these expertise indices are not without their limitations.

We highlighted that the $x$-index may get biased by keywords that are common to multiple disciplines, giving rise to 'overlap bias'. There was a need for a modification of the $x$-index to account for overlapping keywords. To achieve this, we introduced the $x_c$-index, which utilises category-specific keywords over keywords, counting each occurrence of the same keyword in multiple categories as a unique keyword. We believe this modification is sufficient to produce core competency estimates robust to overlap bias.

Unlike the $x$-index, the $x_d$-index had been previously subjected to rigorous testing to assess potential biasing factors (Nandy et al., 2023). These had led to the development of the fractional and field normalised variants of the $x_d$-index. The fractional variant adjusts for multi-institutional collaborative efforts. For instance, often, publications from renowned institutions are perceived as more trustworthy compared to those from less-renowned institutions. This in turn inevitably leads to the former garnering much higher citation counts compared to the latter. Furthermore, publications listing multiple institutions are likely to attract more attention, reaching a wider audience at baseline (immediately post-publication) compared to individual efforts. These aspects may impact citation counts, making comparative analyses between such publications inconsistent. The fractional $x_d$-index overcomes this limitation by utilising the numbers of contributing institutions as weights for the total citations observed for the respective publications.

On the other hand, the field normalised variant aims to adjust for differences in citation norms in different disciplines. Publications from one discipline may receive much higher citations relative to publications from another discipline (Nandy et al., 2023). In traditional analysis, the one receiving higher citation is given more weightage. However, these might be artefacts of the citation norms in those two disciplines. For example, the difference in length of experiments between biomedical sciences and agricultural sciences would lead to differences in citation counts as newer agricultural studies which cite the previous study (i.e., validate previous findings or otherwise) take longer to perform (years) compared to biomedical studies (weeks).

To address this, the field normalised $x_d$-index was introduced, which utilises the discipline-specific mean citation scores to adjust for disciplinary citations norms (Nandy et al., 2024). However, we propose that this field normalised variant may not be unbiased in all situations. Mean citation counts do not always capture the true distribution of citations, primarily when citation counts are not observed to follow a normal distribution. As in the previous example, biomedical sciences is likely to have a left skewed distribution (most works are highly cited), while slower disciplines such as agricultural sciences is likely to have a right skewed distribution (most works observe low citations), resulting in inconsistent means which may introduce further bias into estimates of expertise diversity.

To further overcome differences in field specific citation norms, we introduced the inverse variance weighted (IVW) $x_d$-index which accounts for the differences in variability of citation norms in different fields, thus adjusting for the differences in distributions rather than just the mean. Using means as weights assumes similar distributions across all disciplines of interest which may not always be true. Furthermore, using the inverse of the variances of observed citations will potentially filter out the most consistent disciplines within an institution, adding further insight. However, a limitation of the IVW $x_d$-index is that it may produce biased estimates when disciplines with a low number of publications are included in the analysis, thus requiring stringent quality control and filtering of the dataset.

The best or the most suitable technique for estimating the true variability of citations in different research fields is debatable. One way to achieve this would be to use the list of all publications worldwide. However, this would entail computing variances on entire databases (such as WoS) which is not feasible for most individuals. Another way would be to compute variances on subsets of these databases. For instance, for an institution of interest, variances may be calculated based on the list of publications originating from the same nationality as that of the institution, within a time period. A large enough sample should act as good estimates of worldwide variance. However, there is the potential that these estimates might get biased due to confounding by geography and/or the time period of the generated list.

On the other hand, variances may be estimated solely based on the publications from the institution of interest. Again, a large enough sample would ensure that these variances are representative of the true distributions, albeit rescaled (potentially smaller estimates due to normal sampling error), at least specific to the geography. Moreover, while we have pointed out this issue in the context of the IVW $x_d$-index, it is also likely to affect the computation of means when using the field normalised $x_d$-index. Similar to variances, estimates for field-specific means are likely to be affected by the underlying publication set, geography, and timeline. Further work is required to ascertain any similarities or dissimilarities in mean and variance estimates due to differences in sampling.

Lastly, we introduced a novel index, the overall expertise index or the $x_o$-index. The $x_o$-index may be simply defined as a composite index, comprising the $x$- and $x_d$-indices or, more technically, as the x-weighted $x_d$-index. The $x_o$-index captures both the breadth and depth of research expertise, serving as a measure of overall expertise or strength. Furthermore, the $x_o$-index is unlikely to suffer from the same drawbacks of the $x$- and/or the $x_d$-indices. For instance, the issue of overlap bias is inherently overcome as the $x_o$-index is computed in a category-specific manner. Likewise, the influence of disciplinary differences in distribution of observed citations is negligible as discipline-specific $x$-indices replace citations as weights. However, testing these hypotheses and/or limitations is beyond the scope of this study, but should be visited in future works.

Additionally, we introduced an R package 'xxdi' to help calculate these indices (Das & Nandy, 2026). The open-source software package includes functions to calculate the traditional expertise indices, their original variants, as well as the newly proposed variants and indices. We have structured the functions to be user friendly, even for those without any programming experience. For further guidance, a detailed tutorial is available on GitHub (https://github.com/nilabhrardas/xxdi). It is noteworthy that the formulations of the different weighted rank ratios (i.e., CRR, IVW CRR, and XRR) used in the equations above represent the $h$-index style thresholds, giving the $h$-type expertise indices. Likewise, weighted rank ratios based on the $g$-index may also be calculated to produce the $g$-type expertise indices. All index calculation functions within the 'xxdi' package include a 'type' parameter allowing the user to specify whether to compute the $h$-type or $g$-type expertise index.

## Implications

These expertise indices provide a multi-resolution research strength assessment system, uniquely supporting collaboration planning, diversity expansion, and strategic research portfolio management. The indices capture both the depth and the breadth of research expertise of an institution. This enables a more nuanced understanding of institutional research portfolios. At the institution level, the indices support evidence-based decisions on capacity building, prioritisation of emerging thematic areas, and identification of complementary collaboration partners. At higher aggregation levels, such as regional or national research systems, the nested indices help in identifying clusters of core-competent and emerging institutions (**Table 1**). This, in turn, supports funding allocation, policy design, and long-term ecosystem planning.

Overall, the framework aligns institutional research strategies with broader science and innovation policy objectives, while remaining grounded in transparent and reproducible bibliometric principles. Example usage of the expertise indices is listed in the table below (**Table 1**).

**Table 1.** Strengths and usages of expertise indices.

| Indicator | Recommended Usage | Focus | Key Strength |
|---|---|---|---|
| $x$ | Institutional departments, schools, or discipline-specific research bodies | Fine-grained thematic areas | Core competency |
| $x_d$ | Multidisciplinary and/or discipline-specific institutions or research bodies | Broad thematic areas | Expertise diversity |
| $xx$ and $xx_d$ | Regional domains comprising multiple research bodies (e.g., cities, states, nations) | Institutional diversity clusters | Portfolio planning |
| $x_o$ | Multidisciplinary institutions or research bodies | Breadth and depth of thematic areas | Overall diversity and competency |

# Conclusion

In summary, we revisited the expertise indices, namely the $x$- and the $x_d$-indices and suggested that the traditional expertise indices may not be suitable for use in all situations, and existing bias-aware variants may not be adequate. To address these concerns, we introduced a category adjusted variant of the $x$-index which accounts for overlap bias among keywords, and an inverse variance weighted $x_d$-index which accounts for differences in distributions of acquired citation counts prevalent in different research fields. Additionally, we also introduced another bespoke index, the $x_o$-index, which is a measure of overall research expertise. Lastly, we developed an open-source software package in R dedicated to computing these expertise indices and their variants. The software package, named 'xxdi', is available for download on CRAN (https://doi.org/10.32614/CRAN.package.xxdi) and includes user-friendly functions for those without any programming experience. A detailed tutorial is also available on GitHub (https://github.com/nilabhrardas/xxdi).